# The rise of 212 MAX phase borides, $Ti_2PB_2$, $Zr_2PbB_2$, and $Nb_2AB_2$ [A = P, S]: DFT insights into the physical properties for thermo-mechanical applications

M. A. Ali[1,2,*], M. M. Hossain[1,2], M. M. Uddin[1,2], A. K. M. A. Islam[3,4], S. H. Naqib[2,4,*]

[1]Department of Physics, Chittagong University of Engineering and Technology (CUET), Chattogram-4349, Bangladesh

[2]Advanced Computational Materials Research Laboratory (ACMRL), Department of Physics, Chittagong University of Engineering and Technology (CUET), Chattogram-4349, Bangladesh

[3]Department of Electrical and Electronic Engineering, International Islamic University Chittagong, Kumira, Chattogram-4318, Bangladesh

[4]Department of Physics, University of Rajshahi, Rajshahi-6205, Bangladesh

**Abstract**

An interesting class of ternary metallic borides, known as the 212 MAX phase borides, is the recent advancement of the MAX phase family. This new class exhibits a similar level of metallicity but significantly improved mechanical and thermal properties compared to its traditional carbide counterparts, owing to the existence of a 2D boron layer by forming 2c-2e covalent bonds between the atoms. In this article, results from ab-initio calculations on unexplored $Ti_2PB_2$, $Zr_2PbB_2$, and $Nb_2AB_2$ [A = P, S] are reported wherein $Ti_2PB_2$ along with its 211 boride phase $Ti_2PB$ are predicted for the first time. The stability was confirmed by calculating the formation energy, phonon dispersion curve, and elastic stiffness constants. The obtained elastic constants, elastic moduli, and Vickers hardness values of $Ti_2PB_2$, $Zr_2PbB_2$, and $Nb_2AB_2$ [A = P, S] were found to be significantly larger than those of their counterparts 211 borides and carbides, in a trend similar to other 212 borides. The studied compounds are brittle like most of the MAX and MAB phases. The electronic band structure and density of states revealed the metallic nature of the titled borides. Several thermal parameters were explored, certifying the suitability of $Ti_2PB_2$, $Zr_2PbB_2$, and $Nb_2AB_2$ [A = P, S] compared to their counterparts, and a similar trend was found for the other 212 borides. The obtained results predict that $Ti_2PB_2$, $Zr_2PbB_2$, and $Nb_2AB_2$ [A = P, S] have significant potential for use as efficient thermal barrier coating materials. The response of $Ti_2PB_2$, $Zr_2PbB_2$, and $Nb_2AB_2$ [A = P, S] to the incident photon was studied by computing the dielectric constant (real and imaginary part), refractive index, absorption coefficient, photoconductivity, reflectivity, and energy loss function. The ability to protect from solar heating was revealed from the studied reflectivity spectra. In this work, we have explored the physical basis of the improved thermo-mechanical properties of 212 MAX phase borides compared to their carbide and boride counterparts.

## 1. Introduction

The demand for novel materials with a better demonstration of their performances in various applications is increasing day by day in association with the advancement of technology. One of the best ways to achieve this goal is the prediction and/or synthesis of novel materials in addition to the existing ones that are used in various technological applications. Since the practical use of MAX phase results due to their peculiar physical properties, the prediction, synthesis as well as study of newly synthesized MAX phase materials have received huge interest from both applications as well as basic research points of view [1]. The term MAX represents a layered class of solids in which M is a representative of the transition metal group, A is a representative of the IIIA or IVA group in the periodic table and X is a C/N/B atom [2–8]. Despite the metallic nature of the MAX phase materials, they are also potential candidates for use as an alternative to high-temperature materials because of their ceramic-like characteristics [1,4,5,9,10]. A long list of their potential applications can be found elsewhere [1,5,11–13].

*Corresponding authors: ashrafphy31@cuet.ac.bd; salehnaqib@yahoo.com

However, the main focus for the discovery of a new MAX phase was limited due only to the extension of either the M element or the A element [14–23] or reporting of the MAX phase alloys by the combination of M, M′, and A, A′ atoms; M, M′ = Ti, Zr, Hf, Ta, ….; A, A′ = Al, Ga, Si, Ge, P, … [24–33]. The atom X was recognized as either C/N for a long time before the groundbreaking work of Khazaei et al. [5] where they have proposed B as an X element. Before this proposal, very few attempts were made to extend the MAX phase family by tuning the X elements in comparison with the attempts taken to extend by choosing the M and/or A element(s) for the same. Both the physical and chemical characteristics of B as well as B-containing compounds convey the prospects of the MAX phase borides by replacing C/N with boron [34]. In recent times such systems have been synthesized and are already listed as promising members of the MAX phases. Significant attention has also been paid to these MAX phase borides [6–8,35–43]. The first report on the hypothetical MAX phase borides [$M_2AlB$ (M = Sc, Ti, Cr, Zr, Nb, Mo, Hf, or Ta)] was published by Khazaei et al. [6] where the trend in the electronic structures of the phase stability has been investigated. Gencer et al. [36] reported the electronic and vibrational properties of $Ti_2SiB$. A predictive study of $M_2AlB$ (M = V, Nb, Ta) borides was performed by Surucu et al. [44]. The first synthesis of the $M_2SB$ (M = Zr, Hf, and Nb) borides phase was carried out by Rackl et al. [7,40]. Chakraborty et al. [41] predicted the $V_2AlB$ boride by B substitution in place of C in the $V_2AlC$. A DFT study of $M_2AB$ (M = Ti, Zr, Hf; A = Al, Ga, In) compounds was carried out by G. Surucu [42] in which the structural, electronic, elastic anisotropy, and lattice dynamical properties were considered. We have performed a comprehensive investigation of the synthesized borides $M_2SB$ (M = Zr, Hf, and Nb) in our earlier report [43] where the physical properties of borides were compared with those of carbides. Boron substitutional effect on the carbon site in the $Nb_2SC$ MAX phase was studied by Mitra et al. [37]. Substitution of C and N in place of the B-site of the first synthesized MAX boride $Nb_2SB$ has also been reported [35]. Physical properties of predicted MAX phase borides $Hf_2AB$ (A = Pb, Bi) have been reported by Hossain et al. [39]. Only one MAX phase boride belonging to the 413 sub-class has been investigated by A. Gencer [38] so far.

Miao et al. [8] have predicted MAX phase borides $Hf_2AB$ (A = Bi, Pb) where they have calculated the formation energy and investigated their dynamical stability. Miao et al. [8] have also predicted another class of MAX phase materials crystallizing with different space groups (S.G.: 187, also called sub-space group of conventional MAX phase [8,45]) of the hexagonal system. The traditional MAX phases are crystallized in the S.G.: 194 of the hexagonal system. The study [8] predicted six 212 [$Ti_2InB_2$, $Hf_2AB_2$ (In, Sn), $Zr_2AB_2$ (A = In, Tl, Pb)] MAX phases and two 314 [$Hf_3PB_4$, $Zr_3CdB_4$] MAX phases among which $Ti_2InB_2$ (; No. 187) has already been synthesized [45]. Miao et al. [8] were inspired by the discovery of $Ti_2InB_2$ [45] and other layered ternary borides known as the MAB phases have been investigated [46–48] in recent times. Although the B containing 212 and 314 MAX phases have layered structures that crystallize in a hexagonal system, the atomic arrangement in the cell is different from the conventional MAX phases as shown in Fig. 1. As evident from the figure, there is a 2D layer of B sandwiched in between two Zr layers that significantly contribute in the enhancement of the structural stability and mechanical strength [49].

Li et al. [50] also predicted some MAX phase borides consisting of 212 $Nb_2AB_2$ [A = P, S] and 211 $Nb_2AB$ [A = P, S]. Li et al. [50] have also compared their electronic and transport properties (e.g., the thermal conductivity) with 211 carbides $Nb_2AC$ [A = P, S]. The stability of these phases was reported in terms of formation energy and dynamical stability. These compounds there are predicted to possess unusual thermal properties owing to the acoustic and optical contributions; and the anisotropic nature.

Motivated by the study of Miao et al [8] and Li et al [50], we have tried to explore the 212 $Ti_2PB_2$ and 211 $Ti_2PB$ MAX phase borides by the substitutional method. To do this, at first, we created the structure of $Nb_2PB_2$ and $Nb_2PB$ and calculated their physical properties. After then, we substituted the Nb atoms with Ti in both cases. Recently, the substitutional method has also been used by researchers to predict MAX phase borides [38,41]. Finally, we calculated their formation energy and checked their dynamical and elastic stability.

So far, the physical properties of Ti$_2$InB$_2$ [51,52], Hf$_2$AB$_2$ (A = In, Sn) [53], Zr$_2$AB$_2$ (A = In, Tl) [49], Hf$_3$PB$_4$ [54], and Zr$_3$CdB$_4$ [55] MAX phases have been studied using the density functional theory (DFT) method. For each of the cases, the mechanical properties of B-containing compounds are found to be enhanced remarkably in comparison with their conventional C/N containing 211 MAX phases. The Debye temperature and melting temperature increased for boron-containing 212 phases compared to 211 carbides/nitrides. The minimum thermal conductivity also decreases for the same. The thermal expansion coefficient remains reasonably suitable for coating materials in borides. Thus, the enhanced thermo-mechanical properties of B-containing 212 MAX phases revealed their appropriateness for applications in high-temperature technology compared to the widely used 211 MAX phase carbides. These features of 212 MAX phases are very much motivational and we are interested to study the yet-to-be-investigated stable 212 MAX phase Zr$_2$PbB$_2$. Though Li et al. [50] studied transport properties, electronic band structure, Fermi surface, etc. of Nb$_2$AB$_2$ [A = P, S] and Nb$_2$AB; the mechanical properties and other thermal parameters such as Debye temperature, melting temperature, etc. are not included in their study. Moreover, we have found that Ti$_2$PB$_2$ (212) and Ti$_2$PB (211) MAX borides are energetically, elastically/mechanically, and dynamically stable.

Therefore, in this study, the thermo-mechanical parameters of Ti$_2$PB$_2$, Zr$_2$PbB$_2$, and Nb$_2$AB$_2$ [A = P, S] compounds have been studied employing the DFT method, and the properties are compared with those of other 212 MAX phases as well with their 211 counterpart carbides (Ti$_2$PC, Zr$_2$PbC, Nb$_2$AC [A = P, S]) and borides (Ti$_2$PB, Zr$_2$PbB, Nb$_2$AB [A = P, S]). It is noted that the 211 boride, Zr$_2$PdB, has also been predicted in this study following the same procedure mentioned above.

2. **Computational methodology**

In this study, the physical properties of Ti$_2$PB$_2$, Zr$_2$PbB$_2$, and Nb$_2$AB$_2$ [A = P, S] have been computed by CAmbridge Serial Total Energy Package (CASTEP) code [56,57] via the plane-wave pseudopotential-based DFT method. The generalized gradient approximation (GGA) of the Perdew–Burke–Ernzerhof (PBE) [58] was used to treat the exchange and correlation functions. The electronic orbitals for B - $2s^2\ 2p^1$, Pb- $6s^26p^2$, and Zr - $5s^24p^6\ 4d^2$ were considered for pseudo-atomic calculations. The cutoff energy and *k*-point grids [59] were set to 500 eV and 10 × 10 × 4. The geometry was relaxed by Broyden Fletcher Goldfarb Shanno's (BFGS) technique [60] whereas density mixing was selected for the electronic structure calculations. Furthermore, the self-consistent convergence of the total energy was set to $5 \times 10^{-6}$ eV/atom, and the maximum force per atom was taken as 0.01 eV/Å. A value of $5 \times 10^{-4}$ Å and 0.02 GPa was fixed for the maximum ionic displacement and maximum stress. The bulk modulus (*B*) and shear modulus (*G*) were computed using Hill's approximation [61,62] which is the average value of the upper limit (Voigt [63]) and lower limit (Reuss [64]) of *B*: $[B = (B_V + B_R)/2]$ and *G*: $[G = (G_V + G_R)/2]$. The $B_V$, $B_R$, $G_V$, and $G_R$ can be expressed as follows: $B_V = [2(C_{11} + C_{12}) + C_{33} + 4C_{13}]/9$ ; $B_R = C^2/M$ ; $C^2 = (C_{11} + C_{12})C_{33} - 2C_{13}^2$ ; $M = C_{11} + C_{12} + 2C_{33} - 4C_{13}$ ; $G_V = [M + 12C_{44} + 12C_{66}]/30$ and $G_R = \left(\frac{5}{2}\right)[C^2 C_{44} C_{66}]/[3B_V C_{44} C_{66} + C^2(C_{44} + C_{66})]$ ; $C_{66} = (C_{11} - C_{12})/2$. The Young's modulus (Y) has been estimated using the equation: $Y = 9BG/(3B + G)$ [65,66]. The equation: $v = (3B - Y)/(6B)$ [65,66] was used to compute the Poison's ratio (*v*). The Cauchy pressure (*CP*) is obtained from the stiffness constants; $CP = (C_{12} - C_{44})$.

## 3. Results and discussion

### *3.1 Structural properties and stability*

The structure of the conventional MAX phase is well known, but the structure of the B-containing 212 MAX phase is less familiar. To distinguish between the structures, the unit cell of both 212 ($Zr_2PbB_2$) and 211 ($Zr_2PbC$) phases is presented as a representative in Fig. 1. As seen, there is a clear difference in the atomic arrangements in these structures which is responsible for the different physical behaviors concerning mechanical, electronic, and thermal properties. Both 212 ($Zr_2PbB_2$) and 211 ($Zr_2PbB/Zr_2PbC$) are crystallized in the hexagonal system with different space groups. The 212 MAX phase belongs to the space group $P\bar{6}m2$, (No. 187) [8] while the 211 MAX phase belongs to the space group of P63/mmc (194) [5]. The positions of M, A, and X atoms are different in the 212 and 211 MAX phases. The M (Zr) atoms are positioned at (1/3, 2/3, $z_M$) in the 211 phase, whereas it is at (0.3333, 0.6667, 0.6935) in the 212 phase. The atomic positions of A (Pb) atoms are at (1/3, 2/3, 3/4) and at (0.6667, 0.3333, 0.0), respectively, for the 211 and 212 phases. The most important difference is the positions and contribution of the X atoms in the bulk crystals. The X (C) atoms occupy the corner position (0, 0, 0) in 211 phase while it (B) occupies two positions: (0.6667, 0.3333, 0.5) and (0.0, 0.0, 0.5) forming a 2D layer sandwiched in between the M layers. Within this 2D layer, the B atoms themselves make a strong covalent bond (B-B) that results in a comparatively more stable structure than that of the counterpart 211 MAX phase borides and carbides.

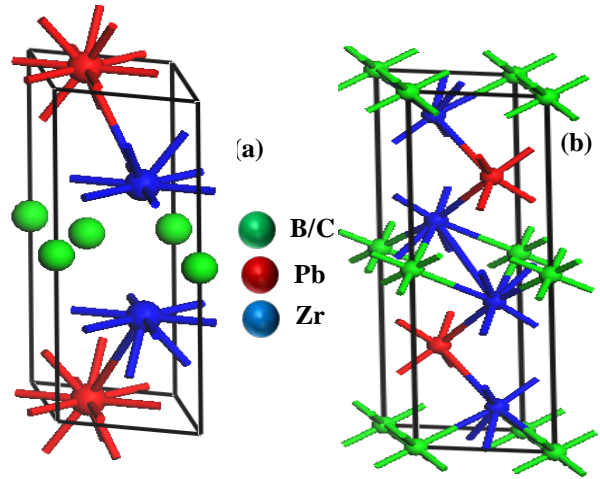

Fig. 1. The unit cells of (a) $Zr_2PbB_2$ and (b) $Zr_2PbB/Zr_2PbC$ MAX compounds.

The unit cell of $Ti_2PB_2$, $Zr_2PbB_2$, and $Nb_2AB_2$ [A = P, S] are optimized to obtain the other physical properties. The cell constants calculated for the optimized structures are given in Table 1. A very good consistency is observed between the values obtained in this study and prior results [8,50]. This reveals the precision of the parameters used for calculations in this work. The lattice constants of 211 borides are presented in Table S1 (in the supplementary document).

**Table 1** - The lattice constants ($a$ and $c$) and $c/a$ ratio of $Ti_2PB_2$, $Zr_2PbB_2$, and $Nb_2AB_2$ [A = P, S] MAX phases.

| Phase | $a$ (Å) | % of deviation | $c$ (Å) | % of deviation | $c/a$ | Reference |
|---|---|---|---|---|---|---|
| $Ti_2PB_2$ | 3.1218 | | 6.5457 | | 2.09 | This work |
| $Zr_2PdB_2$ | 3.2787 | 0.08 | 8.4282 | 0.15 | 2.57 | This work |
| | 3.276 | | 8.415 | | 2.57 | Ref [8] |
| $Nb_2PB_2$ | 3.1929 | 0.53 | 6.6517 | 0.42 | 2.08 | This work |
| | 3.21 | | 6.68 | | 2.08 | [50] |
| $Nb_2SB_2$ | 3.2025 | 0.54 | 6.6548 | 0.52 | 2.07 | This work |
| | 3.22 | | 6.69 | | 2.07 | [50] |

Miao et al [8] have shown that the predicted 212 phases including $Zr_2PbB_2$ are thermodynamically stable by calculating the enthalpy with respect to the known competing phases. They have also calculated the phonon dispersion curves in which no imaginary frequency branch exists. In the case of $Nb_2AB_2$ [A = P, S], Li et al [50] have shown their stability in terms of formation energy and phonon dispersion curves. The $Ti_2PB_2$ has been found to be stable because of the negative formation energy, non-existence of the imaginary frequency in the phonon dispersion curves, and fulfillment of the mechanical stability conditions by the calculated $C_{ij}$s. We have also calculated the phonon dispersion curve of $Zr_2PbB_2$ and $Nb_2AB_2$ [A = P, S] in this study wherein no imaginary frequency branch exists as shown in Fig. 2 (a-d). This confirms the dynamical stability of these borides like the other 212 MAX phase borides. The phonon dispersion of corresponding 211 borides is also presented in Fig. S1 (in the supplementary document). Moreover, the mechanical stabilities of $Ti_2PB_2$, $Zr_2PbB_2$, and $Nb_2AB_2$ [A = P, S] have also been checked in the following section. It is noted that the formation energy of $Ti_2PB_2$ is calculated from the well-known formula used for the prediction of MAX phase materials [67–72]: $E_{form} = \frac{[E_{MAX} - n_M E_M - n_A E_A - n_X E_X]}{n_M + n_A + n_X}$. In this case, we have calculated the energy of all the structures of the elements found in the materials projects [tabulated in Table S2] and computed the formation energy for different combinations. The calculated values of $E_{\text{formation}}$ are found to be in the range from -1.03 to -2.38 eV/atom. Thus, it is expected that the predicted $Ti_2PB_2$ will be chemically stable.

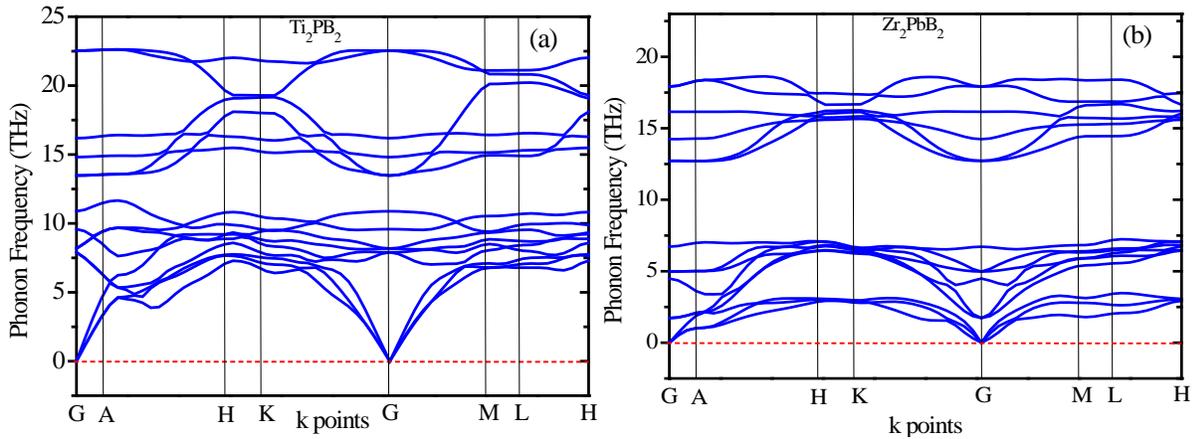

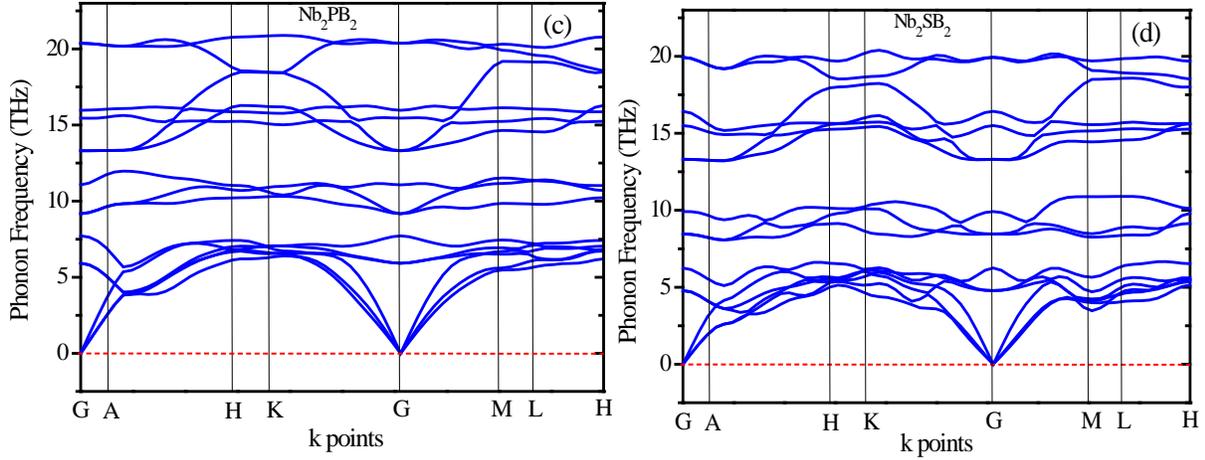

Fig. 2 (a-d). Phonon dispersion curves of Ti$_2$PB$_2$, Zr$_2$PbB$_2$, and Nb$_2$AB$_2$ [A = P, S].

### 3.2.1 *Mechanical properties*

To bring out the figure of merit of the 212 MAX phases in light, it is necessary to discuss the possible relevant parameters for disclosing the mechanical properties. During the practical use of materials, many of them are subjected to applied forces or loads. Thus, knowledge of the related parameters is essential for the proper selection of the materials, otherwise, crack formation, mechanical failure, fracture introduction or structural deformation may take place. One of the key applications of the MAX phases is as structural components at high temperatures [73]. The constants used to unveil the mechanical behavior of solids are stiffness constants, tensile strength, elastic moduli, hardness, brittleness/ductility, fracture toughness, etc. Thus, the aforementioned parameters of Ti$_2$PB$_2$, Zr$_2$PbB$_2$, and Nb$_2$AB$_2$ [A = P, S] have been estimated and discussed [Table 2]. The first step in the study of mechanical properties is the calculation of single crystal elastic constants from which all the other mechanical parameters can be calculated. The obtained single crystal elastic constants of the titled borides using the strain-stress method [51,65,74–76] are tabulated in Table 2 along with those of previously reported 212 MAX phases and their corresponding 211 borides and/or carbides phases. The study of mechanical stability is one of the ways for assessing the stability of solids under static stress used in practical applications. For this case, the stabilities of Ti$_2$PB$_2$, Zr$_2$PbB$_2$, and Nb$_2$AB$_2$ [A = P, S] are checked by the well-known stability conditions [77,78] for hexagonal system as follows: $C_{11} > 0$, $C_{11} > C_{12}$, $C_{44} > 0$, $(C_{11} + C_{12})C_{33} - 2(C_{13})^2 > 0$. The titled borides satisfy the above-stated conditions and hence they are predicted to be mechanically stable.

As can be seen in Table 2, the $C_{11}$ and $C_{33}$ values are not only higher for $Ti_2PB_2$, $Zr_2PbB_2$, and $Nb_2AB_2$ [A = P, S] compared to those of the corresponding 211 boride and carbide phases but also higher for other 212 phases compared to their 211 counterpart carbides. In case of $C_{11}$, the values are 41(33)%, 35(47)%, 25(19)%, and 12(16)% higher for $Ti_2PB_2$, $Zr_2PbB_2$, and $Nb_2AB_2$ [A = P, S], respectively than that of the counterpart borides (carbides). In addition, an increase of 10%, 22%, 11%, 40%, and 28% are also observed from the previously reported phases: $Zr_2InB_2$, $Zr_2TlB_2$, $Hf_2InB_2$, $Hf_2SnB_2$, $Ti_2InB_2$, respectively, compared to their corresponding 211 counterparts. Similarly in case of $C_{33}$, the values are 20(5)%, 38(13)%, 20(19)%, and 23(24)% for $Ti_2PB_2$, $Zr_2PbB_2$, and $Nb_2AB_2$ [A = P, S], respectively than that of counterpart borides (carbides). Besides, an increase of 6%, 21%, 2%, 22%, and 14% is noted for $Zr_2InB_2$, $Zr_2TlB_2$, $Hf_2InB_2$, $Hf_2SnB_2$, and $Ti_2InB_2$, respectively, compared to their corresponding 211 phases. It is well known that the $C_{11}$ and $C_{33}$ measure bonding strength along the *a*- and *c*-axis. Thus, an increase of $C_{11}$ and $C_{33}$ implies the increase of bonding strength along the mentioned directions; consequently, the overall bonding strength is expected to be increased for 212 phases which is also reflected in the values of elastic moduli and hardness values presented in Table 2. As we know, the bulk modulus (*B*) measures the resistance against the uniform volume-changing hydrostatic pressure whereas the shear modulus (*G*) measures the resistance against shape-changing plastic deformation. Comparative stiffness analysis of the solids can be done from the values of Young's modulus (*Y*) which give a measure of the resistance against change in the length. In the case of *B*, the values are found to be increased by 25(8)%, 41(19)%, 16(07)%, and 5(2)% for $Ti_2PB_2$, $Zr_2PbB_2$, and $Nb_2AB_2$ [A = P, S], respectively, compared to their counterpart borides and carbides. Similarly, shear moduli are observed to be increased by 38(34)%, 44(22)%, 23(24)%, and 4(26)%; and Young's moduli are noted to be increased by 35(28)%, 43(21)%, 21(21)%, and 4(22)% for $Ti_2PB_2$, $Zr_2PbB_2$, and $Nb_2AB_2$ [A = P, S], respectively, compared to their counterpart borides and carbides. Moreover, for $Zr_2PbB_2$, $Zr_2InB_2$, $Zr_2TlB_2$, $Hf_2InB_2$, $Hf_2SnB_2$, and $Ti_2InB_2$ phases, the values of B[G(Y)] are increased by 1%[11%(9%)], 17%[18%(17%)], 1%[9%(7%)], 14%[43%(36%)], and 17%[22%(22%)] compared to their corresponding 211 phases $Zr_2InC$, $Zr_2TlC$, $Hf_2InC$, $Hf_2SnC$, and $Ti_2InC$, respectively.

Now, we focus our attention on the hardness parameters, $H_{Chen}$ and $H_{Miao}$, as calculated using the following equations: $H_{Chen} = 2[(\frac{G}{B})^2 G]^{0.585} - 3$ [79] and $H_{Miao} = \frac{(1-2\nu)E}{6(1+\nu)}$ [80]. The calculated values are presented in Table 2. Like the elastic moduli, the hardness parameters of 212 MAX phase borides are also higher than those of 211 MAX phase borides and/or carbides. Moreover, the increase of hardness parameters is also significantly larger for $Ti_2PB_2$, $Zr_2PbB_2$, and $Nb_2PB_2$ while it is small for $Nb_2SB_2$. An increment of 20%(18%), 12%(18%), 16%(15%), 75%(71%), and 21%(26%) for $H_{Chen}$ ($H_{Miao}$) is also reported in the cases of $Zr_2AB_2$ (A = In, Tl), $Hf_2AB_2$ (A = In, Sn) and $Ti_2InB_2$, respectively, compared to those of their corresponding 211 MAX phase carbides. Hence, it is evident that the hardness parameters are significantly enhanced for the 212 phases in comparison with the 211 phases. As discussed so far in this section, the mechanical properties of the 212 MAX phase borides are considerably enhanced compared to their counterpart 211 MAX phase borides and/or carbides. The values of $C_{44}$ for all the studied phases also indicate a significant enhancement as it is most closely related to the hardness of solids [81]. Thus, a reasonable question is, what features bring about these enhancements in the 212 systems? The possible answer is explored in section 3.5.

**Table 2:** The stiffness constants, $C_{ij}$ (GPa), elastic moduli [$B$, $G$, $Y$] (GPa), hardness parameters, Pugh ratios, $G/B$, Poisson ratios, $v$ and Cauchy Pressures, $CP$ (GPa) of $M_2AB_2$ (M =Ti, Zr, Hf, Nb; A = P, S, In, Sn, Tl, Pb) together with those of $M_2AC$ (M =Ti, Zr, Hf, Nb; A = P, S, In, Sn, Tl, Pb) compounds.

| Phase | $C_{11}$ | $C_{12}$ | $C_{13}$ | $C_{33}$ | $C_{44}$ | B | G | Y | $H_{Chen}$ | $H_{Miao}$ | G/B | $v$ | CP | Reference |
|---|---|---|---|---|---|---|---|---|---|---|---|---|---|---|
| $Ti_2PB_2$ | 368 | 90 | 121 | 383 | 184 | 198 | 152 | 361 | 24.74 | 30.79 | 0.77 | 0.19 | -94 | This study |
| $Ti_2PB$ | 261 | 85 | 105 | 320 | 148 | 158 | 110 | 267 | 17.48 | 20.65 | 0.70 | 0.21 | -63 | This study |
| $Ti_2PC$ | 277 | 119 | 131 | 364 | 171 | 184 | 113 | 282 | 14.96 | 19.24 | 0.61 | 0.24 | -52 | This study |
| $Zr_2PbB_2$ | 297 | 57 | 72 | 245 | 76 | 138 | 95 | 232 | 15.55 | 17.73 | 0.69 | 0.22 | -19 | This study |
| $Zr_2PbB$ | 201 | 56 | 48 | 177 | 58 | 98 | 66 | 162 | 11.61 | 12.12 | 0.67 | 0.22 | -02 | This study |
| $Zr_2PbC$ | 220 | 66 | 65 | 216 | 80 | 116 | 78 | 191 | 13.08 | 14.28 | 0.67 | 0.23 | -14 | This study |
| $Nb_2PB_2$ | 438 | 115 | 147 | 476 | 213 | 241 | 179 | 430 | 26.36 | 35.49 | 0.74 | 0.20 | -98 | This study |
| $Nb_2PB$ | 350 | 107 | 141 | 395 | 199 | 207 | 146 | 355 | 21.53 | 27.82 | 0.71 | 0.21 | -92 | This study |
| $Nb_2PC$ | 368 | 123 | 162 | 400 | 194 | 225 | 144 | 355 | 18.72 | 25.24 | 0.64 | 0.23 | -71 | This study |
| $Nb_2SB_2$ | 335 | 90 | 131 | 404 | 155 | 196 | 133 | 326 | 19.20 | 24.58 | 0.68 | 0.22 | -65 | This study |
| $Nb_2SB$ | 326 | 86 | 126 | 328 | 151 | 186 | 128 | 312 | 19.07 | 23.86 | 0.69 | 0.22 | -65 | Ref [43] |
| $Nb_2SC$ | 316 | 108 | 151 | 325 | 124 | 192 | 105 | 267 | 12.02 | 16.22 | 0.55 | 0.27 | -16 | Ref [43] |
| $Zr_2InB_2$ | 315 | 46 | 64 | 263 | 82 | 138 | 105 | 251 | 19.11 | 21.24 | 0.76 | 0.20 | -36 | Ref [49] |
| $Zr_2InC$ | 286 | 62 | 71 | 248 | 83 | 136 | 95 | 231 | 15.87 | 17.94 | 0.70 | 0.22 | -14 | Ref [82] |
| $Zr_2TlB_2$ | 310 | 52 | 61 | 251 | 66 | 135 | 94 | 229 | 15.68 | 17.71 | 0.70 | 0.22 | -21 | Ref[49] |
| $Zr_2TlC$ | 255 | 60 | 52 | 207 | 63 | 115 | 80 | 195 | 13.98 | 15.06 | 0.70 | 0.22 | -03 | Ref [83] |
| $Hf_2InB_2$ | 343 | 61 | 76 | 278 | 94 | 154 | 114 | 274 | 19.46 | 22.56 | 0.74 | 0.20 | -33 | Ref [53] |
| $Hf_2InC$ | 309 | 81 | 80 | 273 | 98 | 152 | 105 | 256 | 16.75 | 19.65 | 0.69 | 0.21 | -17 | Ref[82] |
| $Hf_2SnB_2$ | 353 | 65 | 86 | 306 | 110 | 165 | 124 | 297 | 21.02 | 24.84 | 0.75 | 0.20 | -45 | Ref [53] |
| $Hf_2SnC$ | 251 | 71 | 107 | 238 | 101 | 145 | 87 | 218 | 12.00 | 14.50 | 0.60 | 0.25 | -30 | Ref [84] |
| $Ti_2InB_2$ | 364 | 47 | 58 | 275 | 94 | 147 | 122 | 287 | 23.72 | 26.44 | 0.83 | 0.17 | -47 | Ref [53] |
| $Ti_2InC$ | 284 | 62 | 51 | 242 | 87 | 126 | 100 | 236 | 19.57 | 20.92 | 0.79 | 0.18 | -25 | Ref [52] |

*3.2.2 The brittleness of $Zr_2PB_2$*

Although the MAX phases are metallic, most of them exhibit brittleness like ceramics. In this section, three widely known formalisms are used to predict brittleness/ductility. First of all, the Pugh ratio, ($G/B$) [85] is used to confirm the brittle character of $Ti_2PB_2$, $Zr_2PbB_2$, and $Nb_2AB_2$ [A = P, S] [since $G/B$ < 0.571 for ductile and $G/B$ > 0.571 for brittle solids]. The Poisson's ratio ($v$) also confirmed the brittle character of $Ti_2PB_2$, $Zr_2PbB_2$, and $Nb_2AB_2$ [A = P, S] as $v$ = 0.26 is a critical value to identify the brittle and ductile materials, and $v$ is lower than the critical value for brittle solids, and for ductile solids, the values are higher. In addition, the value of $v$ also indicates the bonding nature within the solids. For example, it is typically low (0.10) for covalent solids and high (0.33) for metallic solids. For the present case, it ($v$) is within the range of 0.19 to 0.23, thus, expecting a mixture of covalent and metallic bonding within the $Ti_2PB_2$, $Zr_2PbB_2$, and $Nb_2AB_2$ [A = P, S] compounds like in other MAX phases. In fact, it agrees well with the existence of the M-X (covalent) bonding and M-A (ionic) bonding within the MAX phase materials. Moreover, Pettifor [86] addressed the Cauchy pressure (*CP*) to determine the chemical bonding and ductile/brittle nature of solids. A negative value of *CP* certifies the covalently bonded brittle solids whereas the positive value indicates the metallic solids. As is evident, $Ti_2PB_2$, $Zr_2PbB_2$, and $Nb_2AB_2$ [A = P, S] belong to the covalently bonded brittle class of solids.

**3.2.3** *Elastic anisotropy*

Complete information regarding the mechanical stability of solids in extreme conditions can be gained from the study of elastic anisotropy. Anisotropy is also related to some other critical processes such as the creation and propagation of microcracks under mechanical stress and anisotropic plastic deformation, etc. Let us look back at the crystal structure again [Fig. 1], the atomic arrangements along the *a(b)*- and *c*-direction are completely different, consequently, the bond strengths along these directions are also different [for example $C_{11} \neq C_{33}$]. These lead to elastic anisotropy in the layered hexagonal system [87]. The knowledge of anisotropy also provides a guideline for the enhancement of the mechanical stability of solids under extreme conditions by supplying direction-dependent elastic constants. Thus, the study of the anisotropic nature of elastic properties is compulsory and we have studied the elastic anisotropy of $Ti_2PB_2$, $Zr_2PbB_2$, and $Nb_2AB_2$ [A = P, S] by calculating some anisotropy indices using different formalisms. The different anisotropic factors for Ti$_2$PB$_2$, Zr$_2$PbB$_2$, and Nb$_2$AB$_2$ [A = P, S] are computed from elastic constants $C_{ij}$ using the following relations [88] and presented in Table 3:

$$A_1 = \frac{1/6\,(C_{11} + C_{12} + 2C_{33} - 4C_{13})}{C_{44}}$$

$$A_2 = \frac{2C_{44}}{C_{11} - C_{12}}$$

$$A_3 = \frac{1/3\,(C_{11} + C_{12} + 2C_{33} - 4C_{13})}{C_{11} - C_{12}}$$

The compounds Ti$_2$PB$_2$, Zr$_2$PbB$_2$, and Nb$_2$AB$_2$ [A = P, S] are anisotropic having the non unit (one) values of $A_i$s because $A_i = 1$ implies the isotropic nature. The linear compressibility ($k$) along the $a$ and $c$-axis are calculated by the equation [89]:

$$\frac{k_c}{k_a} = f = (C_{11} + C_{12} - 2C_{13})/(C_{33} - C_{13})$$

**Table 3:** Anisotropy indices, $A_1$, $A_2$, $A_3$, $k_c/k_a$, and universal anisotropy index $A^U$ of Ti$_2$PB$_2$, Zr$_2$PbB$_2$, and Nb$_2$AB$_2$ [A = P, S] MAX phases along with their counterpart borides and carbides.

| Phases | $A_1$ | $A_2$ | $A_3$ | $k_c/k_a$ | $A^U$ |
|---|---|---|---|---|---|
| Ti$_2$PB$_2$ | 0.67 | 1.32 | 0.89 | 0.82 | 0.14 |
| Ti$_2$PB | 0.64 | 1.68 | 1.07 | 0.63 | 0.33 |
| Ti$_2$PC | 0.58 | 2.16 | 1.27 | 0.58 | 0.67 |
| Zr$_2$PbB$_2$ | 1.22 | 0.63 | 0.77 | 1.21 | 0.45 |
| Zr$_2$PbB | 1.20 | 0.80 | 0.96 | 1.25 | 0.06 |
| Zr$_2$PbC | 0.95 | 1.04 | 0.99 | 1.03 | 0.34 |
| Nb$_2$PB$_2$ | 0.72 | 1.32 | 0.95 | 0.79 | 0.11 |
| Nb$_2$PB | 0.57 | 1.64 | 0.94 | 0.69 | 0.34 |
| Nb$_2$PC | 0.55 | 1.58 | 0.87 | 0.70 | 0.34 |
| Nb$_2$SB$_2$ | 0.76 | 1.27 | 0.96 | 0.60 | 0.11 |
| Nb$_2$SB | 0.62 | 1.26 | 0.78 | 0.79 | 0.15 |
| Nb$_2$SC | 0.63 | 1.19 | 0.75 | 0.70 | 0.16 |

The computed values of $f$ are greater than 1 ($f = 1$ for isotropic materials) for all the phases considered herein which certify the anisotropic nature of the titled compounds.

The universal anisotropy index $A^U$ for Ti$_2$PB$_2$, Zr$_2$PbB$_2$, and Nb$_2$AB$_2$ [A = P, S] is estimated by the following equation [90]: $A^U = 5\frac{G_V}{G_R} + \frac{B_V}{B_R} - 6 \geq 0$, where $B$ and $G$ are bulk and shear modulus; V and R indicate the Voigt and Reuss values. Like other anisotropy indices, $A^U$ for Ti$_2$PB$_2$, Zr$_2$PbB$_2$, and Nb$_2$AB$_2$ [A = P, S] is smaller than one (1) that implies the anisotropic nature of $Ti_2PB_2$, $Zr_2PbB_2$, and $Nb_2AB_2$ [A = P, S]. Since the polycrystalline elastic moduli are calculated from the elastic constants $C_{ij}$, thus, it is obvious that the anisotropic behavior will be exhibited by the moduli as well. Similar results have also been reported for some other 211, 212, 314 MAX phases previously [49,53,54,70,91,92].

*3.3 Electronic band structure (EBS) and density of states (DOS)*

Figs. 3 illustrate the *EBS* of $Ti_2PB_2$, $Zr_2PbB_2$, and $Nb_2AB_2$ [A = P, S] borides in which the electronic paths are shown in the first Brillouin zone. The obtained *EBS* is similar to the metallic solids due to the overlapping of the valence and conduction bands. These are the characteristics of MAX phase materials that exhibit a good combination of metallic and ceramics properties [43,53,93,94].

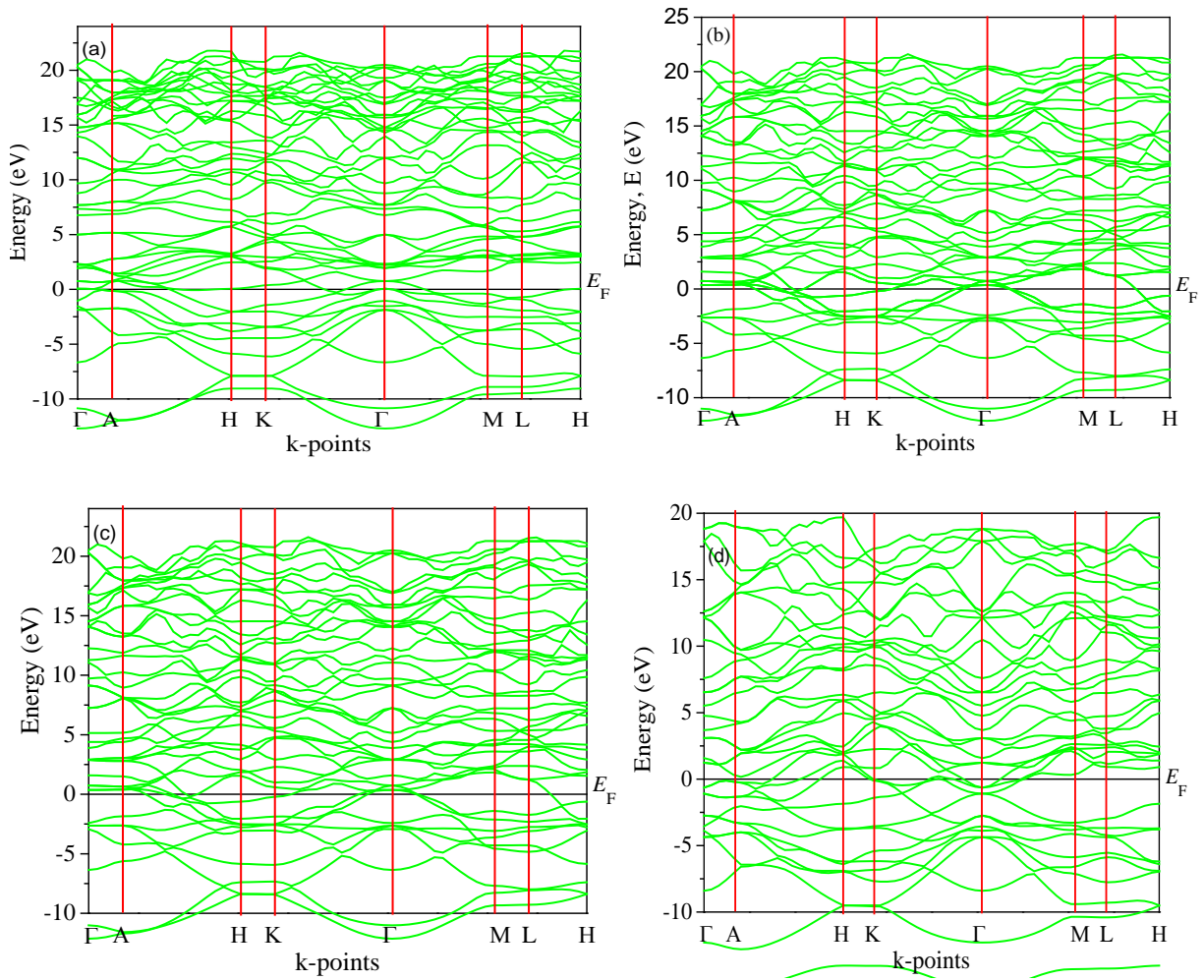

Fig. 3. The electronic band structures of (a) Ti$_2$PB$_2$, (b) Zr$_2$PbB$_2$, (c) Nb$_2$PB$_2$, and (d) Nb$_2$SB$_2$ compounds.

The anisotropic characteristics of the MAX phases are expected due to their layered structure. The EBS also exhibits anisotropy along the basal plane and *c*-direction of the crystal. The paths Γ-A, H-K, and M-L correspond to the *c*-direction while the paths A-H, K-Γ, Γ-M, and L-H are for the basal plane [95]. It is seen that the energy dispersions along the basal plane and the *c*-direction differ significantly, confirming the anisotropic behavior of charge effective masses for these directions, as has been found for other MAX phase nanolaminates [43,49,53].

Figs. 4 (a) displays the partial and total DOS of Zr$_2$PbB$_2$ as a representative. These profiles roughly exhibit the usual characteristics of MAX phase materials. The Fermi level is dominantly contributed by the Zr-*d* electronic states in association with a very small contribution from the B-*p* and Pb-*p* electronic states. The most strong orbital hybridization is observed between B-*p* states and Zr-*d* states that contribute to the formation of strong covalent bonding between them, similar to other MAX phases [12,70,96], specially matched with those of Zr$_2$AB$_2$ (A = In, Tl) [49]. It is worth stating that the Fermi level of Zr$_2$PbB$_2$ resides close to the pseudogap in the TDOS profile. This is an indication of a high level of electronic stability. Similar results are also found for other considered compounds [not shown here]. Fig. 4 (b) shows the total DOS of the titled phases for a comparative understanding of their nature.

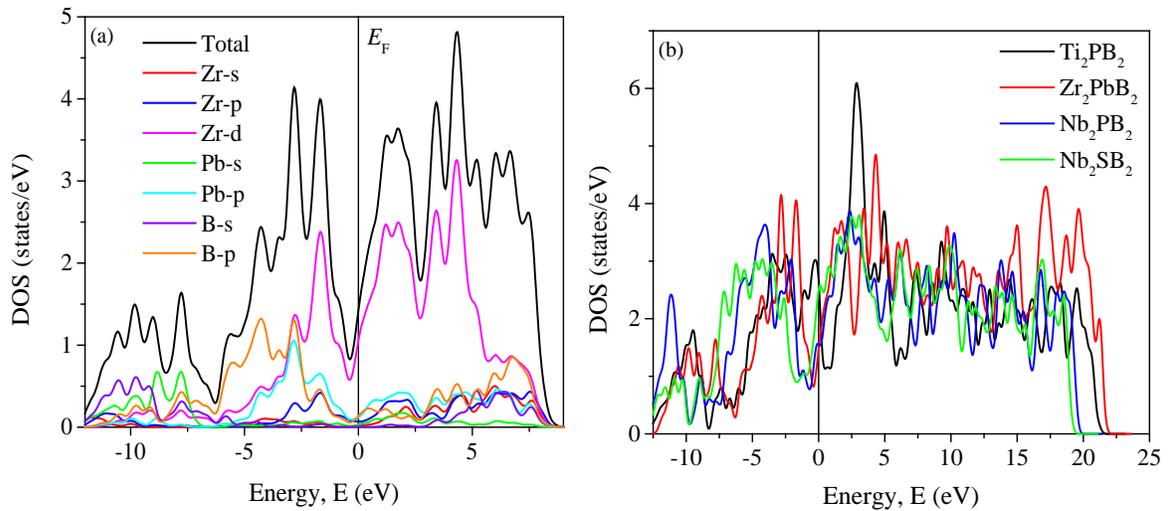

Fig. 4. The total and partial DOS of (a) Zr$_2$PbB$_2$ and (b) total DOS of Ti$_2$PB$_2$, Zr$_2$PbB$_2$, and Nb$_2$AB$_2$ [A = P, S] compounds.

*3.4 Thermal Properties*

Recently, the high-temperature applications of MAX phases have attracted much attention from both research and application points of view. Some of the thermal parameters of particular interest can be used to assess the suitability of materials for high-temperature applications. In this section, the Debye temperature ($\Theta_D$), minimum thermal conductivity ($k_{min}$), thermal expansion coefficient, Grüneisen parameter, and melting temperature of $Ti_2PB_2$, $Zr_2PbB_2$, and $Nb_2AB_2$ [A = P, S] compounds have been calculated to assess their potential for thermal applications.

The elastic properties can be correlated with the thermal properties, for example, phonon dynamics, thermal expansion, thermal conductivity, specific heat, and lattice enthalpy are connected to the Debye temperature. The Debye temperature itself depends on the elastic parameters which determine the sound velocity in a material. The details of the calculations can be found elsewhere [49]. The obtained values of $\Theta_D$ for $Ti_2PB_2$, $Zr_2PbB_2$, and $Nb_2AB_2$ [A = P, S] are 861 K, 434 K, 745 K, and 643 K, much higher than that of the corresponding 211 borides and/or carbides [Table 4]. The $\Theta_D$ values of $Ti_2PB_2$ and $Nb_2AB_2$ [A = P, S] are also higher than those of prior known 212 phases while $\Theta_D$ of $Zr_2PbB_2$ is lower than that of $Zr_2InB_2$, $Hf_2SnB_2$, and $Ti_2InB_2$ but greater than those of $Zr_2TlB_2$ and $Hf_2InB_2$. In reality, the obtained values of $\Theta_D$ for 212 MAX phases are significantly larger compared to those of their B and/or C-containing counterparts belonging to the 211 MAX phases, indicating their higher temperature limits for use because the $\Theta_D$ limits the normal modes of thermal vibrations within the solids. We can quantify the degree of the increment of Debye temperature as 22.3(22.6)%, 26.3(18.6)%, 17.3(18.8)%, and 8.2(13.8)% for $Ti_2PB_2$, $Zr_2PbB_2$, and $Nb_2AB_2$ [A = P, S], respectively, compared to their 211 borides (carbides). The ranking of the so far known 212 MAX phases based on the values of $\Theta_D$ should be as follows: $Ti_2PB_2$ > $Nb_2PB_2$ > $Nb_2SB_2$ > $Ti_2InB_2$ > $Zr_2InB_2$ > $Hf_2SnB_2$ > $Zr_2PbB_2$ > $Zr_2TlB_2$ > $Hf_2InB_2$.

For the assessment of the high-temperature applications, minimum thermal conductivity ($k_{min}$) is one of the important parameters. The thermal conductivity attains a constant value (minimum) at high temperatures known as the minimum thermal conductivity [97]. The lower the value of $k_{min}$, the more suitable the solid for use in thermal barrier coating applications as it conducts less amount of heat which is one of the main criteria for selecting TBC materials. The $k_{min}$ is closely associated with the acoustic wave velocity in the materials and can be calculated by the following model due to Clark [97]: $K_{min} = k_B v_m \left(\frac{M}{n\rho N_A}\right)^{-\frac{2}{3}}$. The values of $k_{min}$ are significantly lower for $Ti_2PB_2$, $Zr_2PbB_2$, and $Nb_2AB_2$ [A = P, S] phases compared to those of the corresponding boride and carbide counterparts as seen in Table 4. As also seen in Table 4 that the $K_{min}$ of $Zr_2PbB_2$ is the lowest and equal to those of $Zr_2InB_2$ and $Hf_2InB_2$, certifying a very low capability of heat conduction at high temperatures. Actually, the $K_{min}$ of 212 phases is lower compared to their 211 counterpart borides and/or carbides, revealing more appropriateness of the 212 phases for the use as TBC materials. However, we can rank [starting from low value] the 212 phases based on the values of $K_{min}$ as follows: $Zr_2PbB_2 = Zr_2InB_2 = Hf_2InB_2 < Hf_2SnB_2 < Zr_2InB_2 < Ti_2InB_2 < Nb_2SB_2 < Nb_2PB_2 < Ti_2PB_2$.

Owing to the close connection of the Grüneisen parameter ($\gamma$) with the specific heat at constant volume, bulk modulus, and TEC, the calculation of $\gamma$ is of scientific interest. It is used to reveal the degree of anharmonic effects present in solids. Thus, $\gamma$ has been determined via the relation with the Poisson's ratio [98]: $\gamma = \frac{3}{2}\frac{(1+\nu)}{(2-3\nu)}$. As seen, the anharmonic effect is low for $Ti_2PB_2$, $Zr_2PbB_2$, and $Nb_2AB_2$ [A = P, S]. The values remain within the predicted limit [0.85 to 3.53] for polycrystalline solids associated with the limiting value of the Poisson's ratio [0.05–0.46] [99].

One of the mandatory information for the selection of materials for applications at high-temperature technology is the melting point ($T_m$) which provides an idea of the temperature limit for usage. Owing to the combined ceramic properties of MAX phases with metallic properties, they have been recognized as suitable systems for high-temperature applications since the early days. Hence, the calculation of $T_m$ for the MAX phase is important and has been estimated in this study using the equation relating elastic constants as follows [100]: $T_m = (3C_{11}+1.5C_{33}+354)$ K.

**Table 4:** Debye temperature, $\Theta_D$, minimum thermal conductivity, $K_{min}$, Grüneisen parameter, $\gamma$, and melting temperature, $T_m$ of $M_2AB_2$ (M =Ti, Zr, Hf, Nb; A = P, S, In, Sn, Tl, Pb) compounds together with those of $M_2AC$ (M =Ti, Zr, Hf, Nb; A =P, S, In, Sn, Tl, Pb) compounds.

| Phase | $\rho$ (g/cm³) | $v_l$ (m/s) | $v_t$ (m/s) | $v_m$ (m/s) | $\Theta_D$ (K) | $K_{min}$ (W/mK) | $\gamma$ | $T_m$ (K) | Reference |
|---|---|---|---|---|---|---|---|---|---|
| $Ti_2PB_2$ | 4.45 | 9480 | 5839 | 6442 | 861 | 1.71 | 1.26 | 2033 | This study |
| $Ti_2PB$ | 4.25 | 8460 | 5083 | 5623 | 704 | 2.43 | 1.35 | 1617 | This study |
| $Ti_2PC$ | 4.61 | 8513 | 4947 | 5489 | 702 | 2.12 | 1.47 | 1731 | This study |
| $Zr_2PbB_2$ | 8.70 | 5514 | 3304 | 3655 | 434 | 0.77 | 1.36 | 1613 | This study |
| $Zr_2PbB$ | 8.25 | 4748 | 2828 | 3131 | 342 | 0.88 | 1.38 | 1338 | This study |

| | | | | | | | | |
|---|---|---|---|---|---|---|---|---|
| Zr$_2$PbC | 9.04 | 4933 | 2937 | 3252 | 366 | 0.97 | 1.41 | 1223 | This study |
| Nb$_2$PB$_2$ | 6.74 | 8435 | 5153 | 5696 | 745 | 1.45 | 1.29 | 2382 | This study |
| Nb$_2$PB | 6.72 | 7726 | 459 | 5151 | 635 | 1.80 | 1.34 | 1997 | This study |
| Nb$_2$PC | 6.98 | 7727 | 4541 | 5033 | 627 | 1.85 | 1.44 | 2058 | This study |
| Nb$_2$SB$_2$ | 6.72 | 7449 | 4446 | 4021 | 643 | 1.39 | 1.37 | 1965 | This study |
| Nb$_2$SB | 6.70 | 7294 | 4366 | 4831 | 594 | 1.14 | 1.36 | 1824 | Ref [43] |
| Nb$_2$SC | 6.31 | 7308 | 4079 | 4542 | 565 | 1.09 | 1.40 | 1790 | Ref [43] |
| Zr$_2$InB$_2$ | 6.87 | 6358 | 3907 | 4312 | 516 | 0.92 | 1.28 | 1693 | Ref [49] |
| Zr$_2$InC | 7.32 | 6004 | 3616 | 3999 | 459 | 1.24 | 1.36 | 1584 | Ref [82] |
| Zr$_2$TlB$_2$ | 8.71 | 5466 | 3284 | 3633 | 433 | 0.77 | 1.36 | 1660 | Ref [49] |
| Zr$_2$TlC | 8.92 | 4989 | 2993 | 3311 | 372 | 0.89 | 1.36 | 1430 | Ref [83] |
| Hf$_2$InB$_2$ | 10.86 | 5309 | 3240 | 3578 | 431 | 0.77 | 1.29 | 1800 | Ref [53] |
| Hf$_2$InC | 11.67 | 5004 | 2999 | 3319 | 383 | 1.04 | 1.32 | 1691 | Ref [82] |
| Hf$_2$SnB$_2$ | 11.08 | 5459 | 3344 | 3692 | 447 | 0.80 | 1.28 | 1872 | Ref [53] |
| Hf$_2$SnC | 12.06 | 5121 | 3050 | 3376 | 393 | 1.07 | 1.49 | 1746 | Ref [101] |
| Ti$_2$InB$_2$ | 05.90 | 7241 | 4545 | 5004 | 633 | 1.19 | 1.19 | 1858 | Ref [53] |
| | 05.91 | 7168 | 4487 | 4942 | 621 | 1.23 | 1.20 | 1833 | Ref [52] |
| Ti$_2$InC | 06.08 | 6531 | 4055 | 4471 | 534 | 1.00 | 1.23 | 1569 | Ref [52] |

Like the other 212 MAX phases, the $T_m$ of Ti$_2$PB$_2$, Zr$_2$PbB$_2$, and Nb$_2$AB$_2$ [A = P, S] values are also larger than their 211 corresponding counterpart borides and/or carbides. On the other hand, $T_m$ is the lowest [1613 K] for Zr$_2$PbB$_2$ in comparison with other 212 MAX phase borides while it is maximum for Nb$_2$PB$_2$ [2382 K] (Table 4). In addition, the used formula to calculate the $T_m$ contains $C_{11}$ and $C_{33}$, which are the measures of the stiffness along the $a$- and $c$-axis. Thus, a direct relationship between the $T_m$ and Young's modulus [a measure of the stiffness of a polycrystalline solid] is expected which is seen in Table 2 and Table 4. The maximum value of $Y$ is obtained for Nb$_2$PB$_2$, following the ranking as follows: Nb$_2$PB$_2$ > Ti$_2$PB$_2$ > Nb$_2$SB$_2$ > Hf$_2$SnB$_2$ > Ti$_2$InB$_2$ > Hf$_2$InB$_2$ > Zr$_2$InB$_2$ > Zr$_2$TlB$_2$ > Zr$_2$PbB$_2$. This ranking also applies to the melting temperature $T_m$ as seen in Table 4. It has been suggested previously that $T_m$ will be higher for the solids with higher Young's modulus and vice versa [102]. Knowledge regarding the decomposition temperature ($T_d$) is also a prerequisite regardless of the value of $T_m$. The MAX phases have a tendency to decompose at a lower temperature than that of $T_m$. Unfortunately, no information regarding $T_d$ of the titled borides as well as other 212 phases is present at this moment. An earlier report provided by Cue et al. [103] might be helpful regarding this issue. They have estimated $T_d$ and $T_m$ of some MAX phases and shown that $T_d$ is lower than $T_m$, for some cases $T_d$ is close to $T_m$ [54], thus, one may expect the same for Ti$_2$PB$_2$, Zr$_2$PbB$_2$, and Nb$_2$AB$_2$ [A = P, S] phases as well.

## 3.5 Why are the thermo-mechanical properties for 212 phases enhanced?

From the analysis of the results obtained so far, it is obvious that the thermo-mechanical properties are significantly upgraded when we go from the 211 MAX phase carbides to the 212 MAX phase borides. Why the properties are get enhanced for the 212 phases? The answer lies in the structure. Though both the 212 and 211 phases belong to the MAX family, their structures are not the same. The 212 phases are crystallized in the space group P-6m2, (No. 187) [8,45], while the 211 phases are crystallized in the space group P6$_3$/mmc (No.194) [5,104] as shown in Fig. 1 (a and b). The bonding natures within these two structures are also different. For the 212 phases, there is a very strong covalent bonding among the B atoms by forming a 2D layer and contributing to the 2c-2e type of bonding as shown in Figs. 5 (a and b). In addition, there is also strong covalent bonding between M (Zr atoms for Zr$_2$PbB$_2$) and X (B atoms for Zr$_2$PbB$_2$) similar to the 211 phases (Zr and B/C are M and X atoms for Zr$_2$PbB/Zr$_2$PbC). In addition, there is comparatively weak M – A bonding (Pb is the A atom in the case of Zr$_2$PbB/Zr$_2$PbC). There is no covalent bond among the C atoms in the 211 phase, although strong covalent bonds between M- X atoms are formed as well as the M-A bonds. Owing to the existence of the very strong B-B covalent bonding within the 2D layer of B atoms, the structure becomes more stable, and the overall cohesive energy is enhanced significantly for the 212 phases in comparison with the 211 phases. It is noted that, for 211 MAX phase borides, B-B bonding does not exist [6,7,37,39,43,105]. Our statements will be further clarified by the calculation of the Mulliken population analysis given in Table 5. As seen in Table 5, the charge is transferred from Zr to B (0.54/0.55) and Pb (0.01). The charge transfer mechanism confirms the existence of ionic character as observed for the B/C-containing MAX phase Zr$_2$PbB/Zr$_2$PbC. In this boride/carbide, the charge is transferred from Zr to B (0.77) and Pb (0.00) [Table 5]. Table 5 also shows the bond overlap population (BOP) analysis for both the compounds Zr$_2$PbB$_2$ and Zr$_2$PbB. The value of BOP indicates the nature of bonding/anti-bonding states within the solids depending on the positive/negative values. A negative value of BOP stands for anti-bonding whereas a positive value for bonding states. It is clear from Table 5 that no anti-bonding exists within the considered MAX phases. In addition, the value of BOP is also an indicator of bonding strength, the higher the BOP, the stronger the bonding. As evident from Table 5, the B-B bonding exhibits a much higher BOP value, confirming the very strong covalent bonding formed in the 2c-2e channel. The M (Zr) - X (B) bonding in the Zr$_2$PbB is stronger than that of Zr$_2$PbB$_2$ with the

BOP values of 1.25 and 0.24/0.20, respectively. Due to the existence of very strong B-B bonding, the overall bonding strength, as well as mechanical parameters and hardness values, are higher for $Zr_2PbB_2$ compared to $Zr_2PbB$. The results for the other titled compounds are also presented in Table 5. Similar results are also reported for other known 212 phases [49,51,53].

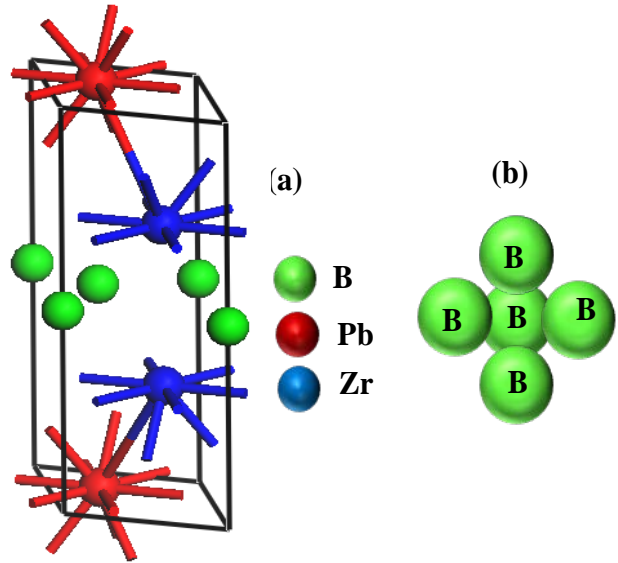

Fig. 5. Formation of B-B covalent bonding within $Zr_2PbB_2$.

Now, let us discuss the variation of the mechanical properties [Table 2] among the titled borides in terms of the BOP values as presented in Table 5. We can rank the titled borides based on the values of hardness parameters [Table 2] ($H_{chen}$, $H_{miao}$, and $C_{44}$) as follows: $Nb_2PB_2$ > $Ti_2PB_2$ > $Nb_2SB_2$ > $Zr_2PbB_2$. The BOP of B-B bonds is 2.22, 2.47, 2.07, and 2.12 for $Nb_2PB_2$, $Ti_2PB_2$, $Nb_2SB_2$, and $Zr_2PbB_2$, respectively. According to these values, the ranking of borides can be re-ordered. But, for $Nb_2PB_2$, the BOPs of Nb-B are (0.14 and 0.19) higher compared to those of Ti-B (0.05 and 0.12), and the BOP of Nb-P and Ti-P is the same for both compounds. Thus, considering the cumulative contribution from all bonds (BOP), the overall bonding strength is expected to be higher for $Nb_2PB_2$ compared to that for $Ti_2PB_2$ which is reflected in values of the elastic constants, elastic moduli, and hardness parameters. Again for $Nb_2SB_2$ and $Zr_2PbB_2$, the BOP of the B-B bond is 2.07 and 2.12, respectively. The BOP of Nb-B (0.20 and 0.19) and Zr-B (0.24 and 0.20) are also comparable. But, there is no BOP for Zr-Pb bonding within the considered threshold length (5 Å) whereas the BOP for Nb-S is 0.74. This makes the difference in the overall bonding strength as reflected from values of elastic constants, elastic moduli, and hardness parameters. It is also noted that no BOP of Zr-Pb is found for 211 the $Zr_2PbB$ compound [Table 5].

**Table 5:** Mulliken atomic and bond overlap populations of $Ti_2PB_2$, $Zr_2PbB_2$, and $Nb_2AB_2$ [A = P, S] and $Ti_2PB$, $Zr_2PbB$, and $Nb_2AB$ [A = P, S] MAX compounds.

| Compound | Atoms | Atomic population | | | | | Bond overlap population | | | |
| --- | --- | --- | --- | --- | --- | --- | --- | --- | --- | --- |
| | | s | p | d | Total | Charge (e) | Bond | Bond number $n^\mu$ | Bond length $d^\mu$ (Å) | Bond population $P^\mu$ |
| $Ti_2PB_2$ | B | 0.95 | 2.62 | 0.00 | 3.56 | -0.56 | B-B | 1 | 1.80242 | 2.47 |
| | B | 0.95 | 2.61 | 0.00 | 3.57 | -0.57 | Ti-B | 2 | 2.39781 | 0.12 |
| | Ti | 2.10 | 6.51 | 2.79 | 11.39 | 0.61 | Ti-B | 2 | 2.39781 | 0.05 |
| | P | 1.58 | 3.50 | 2.79 | 5.08 | 0.08 | Ti-P | 2 | 2.47181 | 1.00 |
| $Ti_2PB$ | B | 1.21 | 2.42 | 0.00 | 3.64 | -0.64 | Ti-B | 4 | 2.27017 | 1.10 |
| | Ti | 2.21 | 6.67 | 2.73 | 11.62 | 0.38 | Ti-P | 4 | 2.49843 | 0.96 |
| | P | 1.61 | 3.52 | 0.00 | 5.13 | -0.13 | | | | |
| $Zr_2PbB_2$ | B | 0.98 | 2.58 | 0.00 | 3.55 | -0.55 | B-B | 1 | 1.89298 | 2.12 |
| | B | 0.97 | 2.57 | 0.00 | 3.54 | -0.54 | Zr-B | 2 | 2.51753 | 0.24 |
| | Zr | 2.15 | 6.46 | 2.84 | 11.45 | 0.55 | Zr-B | 2 | 2.51753 | 0.20 |
| | Pb | 1.42 | 2.55 | 10.04 | 14.01 | -0.01 | | | | |
| $Zr_2PbB$ | B | 1.18 | 2.59 | 0.00 | 3.77 | -0.77 | Zr-B | 4 | 2.38633 | 1.25 |
| | Zr | 2.28 | 6.61 | 2.72 | 11.62 | 0.38 | | | | |
| | Pb | 1.41 | 2.55 | 10.03 | 14.00 | 00.0 | | | | |
| $Nb_2PB_2$ | B | 0.92 | 2.57 | 0.00 | 3.49 | -0.49 | B-B | 1 | 1.84346 | 2.22 |
| | B | 0.95 | 2.56 | 0.00 | 3.51 | -0.51 | Nb-B | 2 | 2.43378 | 0.19 |
| | Nb | 2.15 | 6.28 | 4.07 | 15.51 | 0.49 | Nb-B | 2 | 2.43378 | 0.14 |
| | P | 1.55 | 3.43 | 0.00 | 4.99 | 0.01 | Nb-P | 2 | 2.53279 | 1.00 |
| $Nb_2PB$ | B | 1.14 | 2.46 | 0.00 | 3.60 | -0.60 | Nb-B | 4 | 2.29051 | 1.15 |
| | Nb | 2.25 | 6.41 | 4.02 | 12.68 | 0.32 | Nb-P | 4 | 2.56130 | 1.06 |
| | P | 1.59 | 3.46 | 0.00 | 5.04 | -0.04 | | | | |
| $Nb_2SB_2$ | B | 0.92 | 2.56 | 0.00 | 3.47 | -0.47 | B-B | 1 | 1.84897 | 2.07 |
| | B | 0.93 | 2.58 | 0.00 | 3.51 | -0.51 | Nb-B | 2 | 2.40337 | 0.20 |
| | Nb | 2.14 | 6.27 | 4.04 | 12.45 | 0.55 | Nb-B | 2 | 2.40337 | 0.19 |
| | S | 1.76 | 4.37 | 0.00 | 6.12 | -0.12 | Nb-S | 2 | 2.57488 | 0.74 |
| $Nb_2SB$ | B | 1.11 | 2.51 | 0.00 | 3.62 | -0.62 | Nb-B | 4 | 2.26543 | 1.12 |
| | S | 1.77 | 4.40 | 0.00 | 6.17 | -0.17 | Nb-S | 4 | 2.59817 | 0.79 |
| | Nb | 2.24 | 6.37 | 3.99 | 12.61 | 0.39 | | | | |

### 3.6 Optical properties

The technologically important optical constants of $Ti_2PB_2$, $Zr_2PbB_2$, and $Nb_2AB_2$ [A = P, S] MAX phase compounds are estimated and shown in Fig. 6 in the energy range up to 25 eV. The low energy part of the imaginary part of the dielectric function is corrected by introducing a plasma frequency of 3 eV and damping of 0.5 eV owing to the metallic nature of the titled compounds. Besides, we have used a Gaussian smearing 0.5 eV.

The real part of the dielectric function, $\varepsilon_1$ is shown in Fig. 6 (a) exhibiting metallic behavior. Usually, the $\varepsilon_1$ exhibit a large negative value in very low energy region for metallic systems. Thus, $\varepsilon_1$ spectra of $Ti_2PB_2$, $Zr_2PbB_2$, and $Nb_2AB_2$ [A = P, S] agree well with the non-existence of the energy band gap in the electronic band structure. The imaginary part of dielectric function $\varepsilon_2$ is depicted in Fig. 6 (b) wherein two small peaks are observed for both polarization directions of the electric field vector. The peaks in the $\varepsilon_2$ certify that $Ti_2PB_2$, $Zr_2PbB_2$, and $Nb_2AB_2$ [A = P, S] compounds are metal, and the $E_F$ [Fig. 3] is located at the Ti-$3d$ states that results in the free electron behavior, consequently, the photon-induced transition between electronic states is possible within the conduction band.

The refractive index ($n$) and extinction coefficient ($k$) are depicted in Fig. 6 (c) and (d), respectively. The $n(0)$ is found to be 6.4, 7.8, 7.1 and 18.6 for $Ti_2PB_2$, $Zr_2PbB_2$, and $Nb_2AB_2$ [A = P, S], respectively. The $k$ for $Ti_2PB_2$, $Zr_2PbB_2$, and $Nb_2AB_2$ [A = P, S] rise progressively in the IR region to its highest values and then decline slowly in the visible and UV regions [Fig. 6(d)]. The decline of $k$ follows the decline of $\varepsilon_2$, as seen in Fig. 6 (b) and Fig. 6 (d).

The absorption coefficient ($\alpha$) of $Ti_2PB_2$, $Zr_2PbB_2$, and $Nb_2AB_2$ [A = P, S] is presented in Fig 6 (e). As seen in the figure, it rises almost steadily to attain its maximum value in the UV region. Moreover, $\alpha$ starts rising from the zero photon energy, indicating once again the metallic nature of the compounds under investigation. In addition, it is found that $Ti_2PB_2$, $Zr_2PbB_2$, and $Nb_2AB_2$ [A = P, S] can be used as good absorbing materials in the visible and UV regions owing to the high values of their absorption coefficients. Fig. 6 (f) displays the photoconductivity ($\sigma$) of $Ti_2PB_2$, $Zr_2PbB_2$, and $Nb_2AB_2$ [A = P, S] as a function of photon energy. The $\sigma$ measures the change in the electrical conductivity of the material when it is subjected to photon irradiation. Like $\alpha$, the $\sigma$ spectrum also certifies that $Ti_2PB_2$, $Zr_2PbB_2$, and $Nb_2AB_2$ [A = P, S] are metallic compounds.

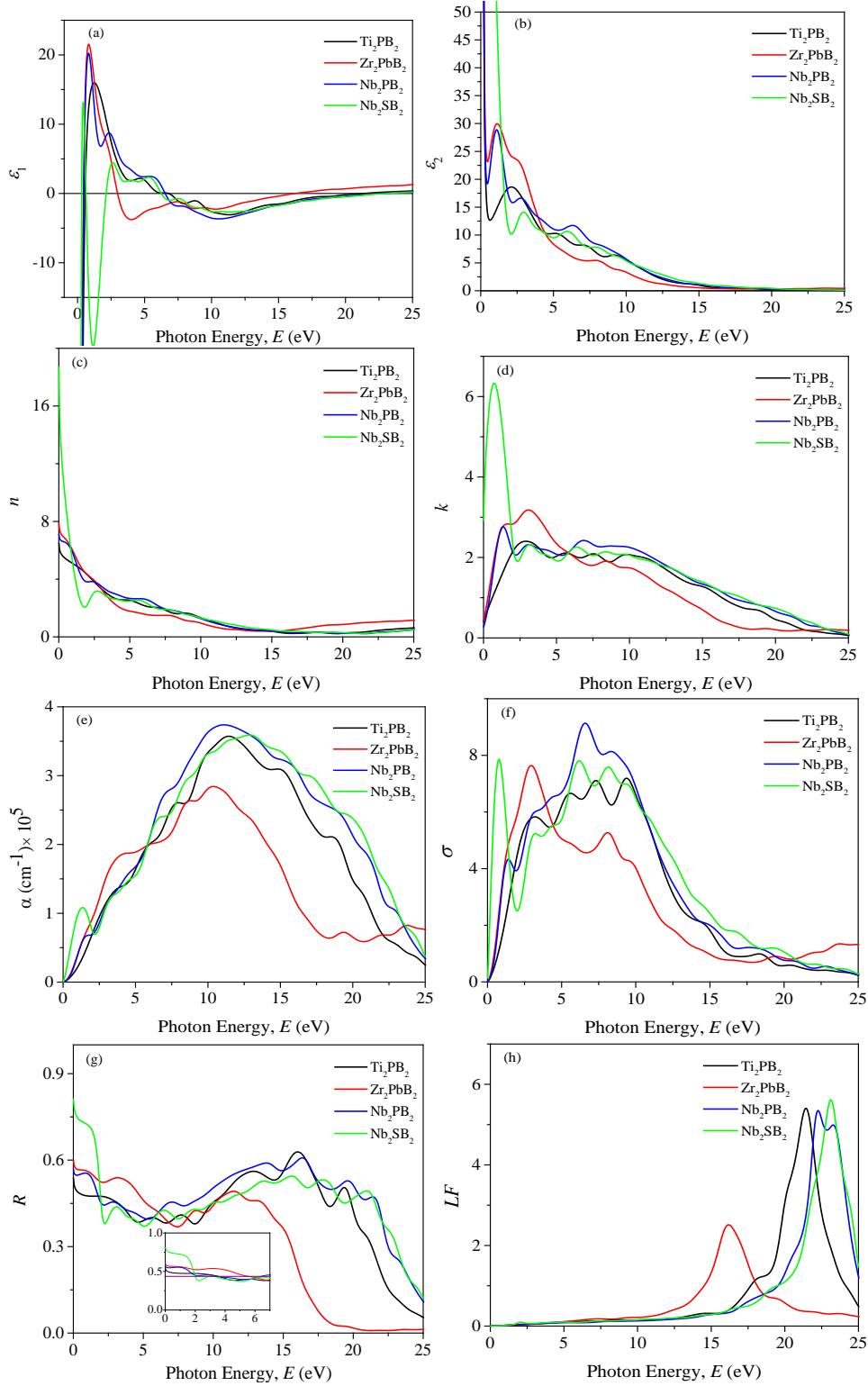

Fig. 6 (a) real part, $\varepsilon_1$, and (b) imaginary part of dielectric function, $\varepsilon_2$, (c) refractive index, $n$, (d) extinction coefficient, $k$, (e) absorption coefficient, $\alpha$, (f) photoconductivity, $\sigma$, (g) reflectivity, $R$ and (h) lossfunction, $L$ as a function of electromagnetic radiation energy.

One of the most important applications of MAX phase materials is as coating materials to reduce solar heating. The potential for this application can be predicted by the study of reflectivity. The reflectivity $R(\omega)$ of $Ti_2PB_2$, $Zr_2PbB_2$, and $Nb_2AB_2$ [A = P, S] is presented Fig. 6 (g). According to Li *et al.* [106,107] the MAX phase materials which have reflectivity greater than 44% are suitable for use as coating materials. As evident from Fig. 6 (g), the $R(\omega)$ remains always greater than 44% up to ~12 eV for all the considered phases. Though the reflectivity at low energy is highest for $Nb_2SB_2$, the overall performance of $Zr_2PbB_2$ is the best for use as a cover to protect from solar heating. Noted here that the reflectivity spectrum with ~ 53% in the IR, visible, and near UV region (0-5 eV) was almost constant for $Zr_2PbB_2$. Fig. 6 (h) shows the energy loss function [$L(\omega)$] spectra of $Ti_2PB_2$, $Zr_2PbB_2$, and $Nb_2AB_2$ [A = P, S] compounds, in which the peaks correspond to plasma frequencies ($\omega_p$). At this particular frequency the $R(\omega)$ exhibits falling a tail, the absorption coefficient falls rapidly and $\varepsilon_1$ crosses zero from the negative side. It is the characteristic frequency above which the materials behave as transparent to the incident electromagnetic wave.

4. **Conclusions**

The thermo-mechanical and optical properties of $Ti_2PB_2$, $Zr_2PbB_2$, and $Nb_2AB_2$ [A = P, S] are explored via DFT calculations with the intention to shed light on the enhanced thermo-mechanical properties of recently added 212 MAX phase borides in comparison with their 211 boride and carbide counterparts. The titled MAX phases are chemically, dynamically, and mechanically stable. The values of elastic constants and moduli of 212 phases are higher than those of the 211 phase borides and/or carbides. The hardness parameters are also higher for the same. The usual brittleness of the MAX phase is also exhibited by the titled borides. Due to the different atomic arrangements along the *a*- and *c*-direction, the elastic properties are anisotropic. The metallic character of the studied phases is confirmed by the EBS and DOS. Like the mechanical properties, the thermal parameters are also enhanced significantly for 212 phases compared to their 211 counterparts. For $\Theta_D$, the values are increased by 22.3(22.6)%, 26.3(18.6)%, 17.3(18.8)%, and 8.2(13.8)% for $Ti_2PB_2$, $Zr_2PbB_2$, and $Nb_2AB_2$ [A = P, S], respectively, compared to the corresponding 211 borides (carbides). The melting temperature is also noted to be increased by 26(17)%, 20(32)%, 19(16)%, and 08(10)%. The study of thermo-mechanical properties reveals the more appropriateness of the 212 phase borides compared to their 211 counterparts MAX phases for applications. The existence of very strong B-B bonding contributes to the enhancement of the thermo-mechanical properties of the 212 compounds. The important results concerning the optical constants are found in accord with the electronic band structure results. The dielectric constant, absorption coefficient, and photoconductivity spectra reconfirm the metallic nature of the titled MAX phases. The reflectivity spectra show the suitability of $Ti_2PB_2$, $Zr_2PbB_2$, and $Nb_2AB_2$ [A = P, S] phases for use as shielding materials to protect from solar radiation.

To conclude, we hope, the results presented in this paper will stimulate further research on the 212 MAX phases which exhibit superior characteristics for applications in the thermo-mechanical sectors compared to many other conventional MAX phase nanolaminates.

**Conflicts of interest**

There are no conflicts of interest to declare.

**Acknowledgment**

The authors are grateful to the Department of Physics, Chittagong University of Engineering & Technology (CUET), Chattogram-4349, Bangladesh, for providing the computing facility for this work. A part of this work has been done at the ACMRL, Department of Physics, CUET, Bangladesh, where the facility is supported by the research grant from The World Academy of Sciences [No. 21-378 RG/PHYS/AS_G -FR3240319526].

Supplementary Information

# The rise of 212 MAX phase borides, Ti$_2$PB$_2$, Zr$_2$PbB$_2$, and Nb$_2$AB$_2$ [A = P, S]: DFT insights into the physical properties for thermo-mechanical applications

M. A. Ali[1,2,*], M. M. Hossain[1,2], M. M. Uddin[1,2], A. K. M. A. Islam[3,4], S. H. Naqib[2,4,*]

[1]Department of Physics, Chittagong University of Engineering and Technology (CUET), Chattogram-4349, Bangladesh
[2]Advanced Computational Materials Research Laboratory (ACMRL), Department of Physics, Chittagong University of Engineering and Technology (CUET), Chattogram-4349, Bangladesh
[3]Department of Electrical and Electronic Engineering, International Islamic University Chittagong, Kumira, Chattogram-4318, Bangladesh
[4]Department of Physics, University of Rajshahi, Rajshahi-6205, Bangladesh

**Table S1 -** The lattice constants (*a* and *c*) and *c/a* ratio of 211 MAX phase borides.

| Phase | a (Å) | c (Å) | c/a | Reference |
|---|---|---|---|---|
| Ti$_2$PB | 3.2674 | 11.6044 |  | This study |
| Zr$_2$PdB | 3.5021 | 15.1729 |  | This study |
| Nb$_2$PB | 3.3403 | 11.6853 |  | This study |
| Nb$_2$SB | 3.5383 | 11.6043 |  | [1] |

**Table S2** Different structures of Ti$_2$PB$_2$ composing elements.

| Phase ID of Ti | Energy of Ti (eV) | Phase ID of P | Energy of P (eV) |
|---|---|---|---|
| mp-72 | -4809.31 | mp-53 | -179.58 |
| mp-46 | -3206.1 | mp-118 | -4309.48 |
| mp-6985 | -1602.97 | mp-130 | -359.271 |
| mp-73 | -1603.02 | mp-157 | -718.668 |
|  |  | mp-7245 | -359.269 |
| Phase ID of B | Energy of B (eV) | mp-12883 | -1436.49 |
| mp-160 | -926.397 | mp-674158 | -174.21 |
| mp-1193675 | -2160.86 | mp-1014013 | -718.626 |
| mp-22046 | -3854.98 | mp-1094075 | -1796.33 |
| mp-570316 | -3697.84 | mp-1179990 | -357.596 |
| mp-570602 | -3848.55 | mp-1198724 | -7547.41 |
| mp-632401 | -923.045 |  |  |
| mp-1202723 | -922.895 | Phase | Energy (eV) |
| mp-1055985 | -74.4371 | Ti$_2$PB$_2$ | -3545.42 |

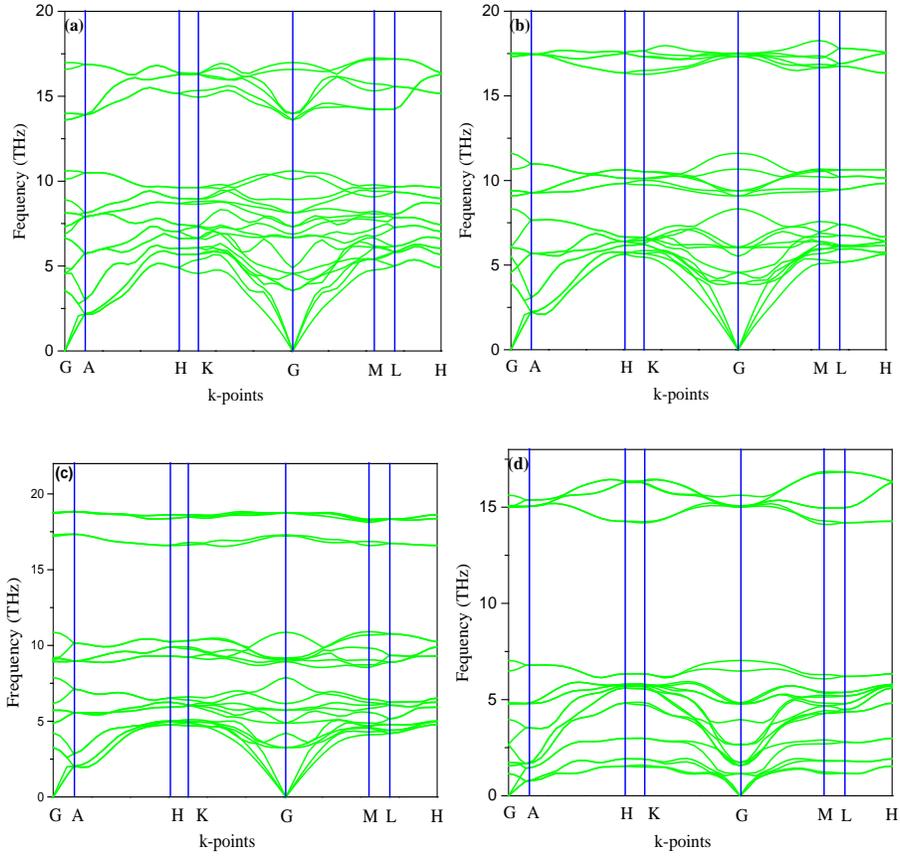

Fig. S1 Phonon dispersion curves of (a) Ti$_2$PB, (b) Zr$_2$PbB, (c) Nb$_2$PB, and (d) Nb$_2$PB MAX phase compounds.

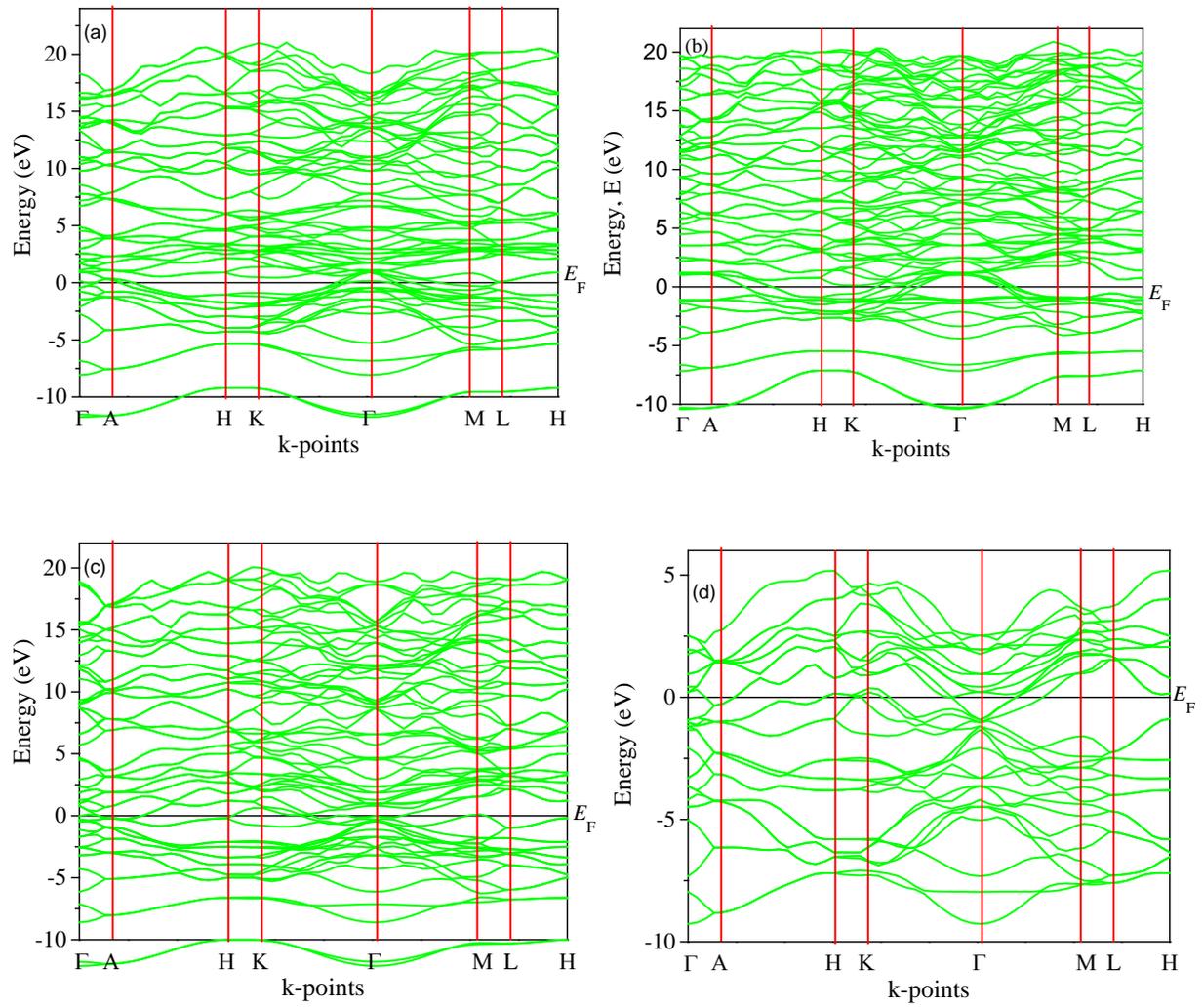

Fig. S2. The electronic band structures of (a) $Ti_2PB$, (b) $Zr_2PbB$, (c) $Nb_2PB$, and (d) $Nb_2SB$ MAX phase compounds.

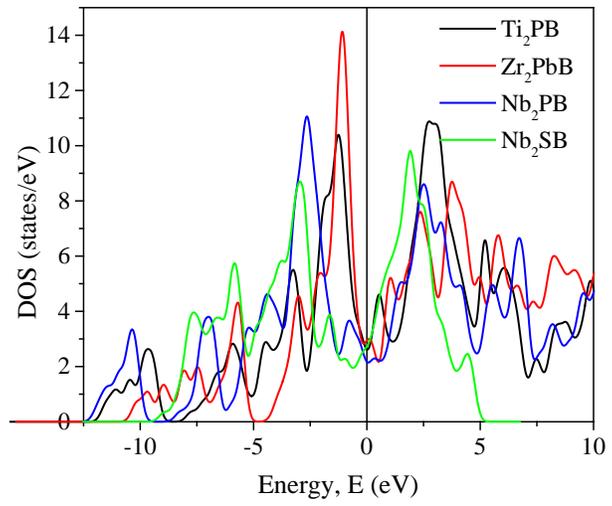

Fig. S3. The total DOS of Ti$_2$PB, Zr$_2$PbB, and Nb$_2$AB [A = P, S] MAX phase compounds.